\documentclass[aps,twocolumn,superscriptaddress,showpacs]{revtex4}
\usepackage{graphicx}

\begin{document}

\title{Current reversal with type-I intermittency in deterministic inertia ratchets}

\author{Woo-Sik Son} \email{dawnmail@physics3.sogang.ac.kr} 

\affiliation{Department of Physics, Sogang University, Seoul 121-742, 
        Korea}

\author{Inbo Kim} \email{ibkim@physics3.sogang.ac.kr}

\affiliation{Department of Physics, Sogang University, Seoul 121-742,
        Korea}

\author{Young-Jai Park} \email{yjpark@sogang.ac.kr}

\affiliation{Department of Physics, Sogang University, Seoul 121-742,
        Korea}

\author{Chil-Min Kim} \email{chmkim@mail.paichai.ac.kr}

\affiliation{National Creative Research Initiative Center for
  Controlling Optical Chaos, \\Department of Physics, Paichai
  University, Taejon 302-735, Korea}

\date{\today}

\begin{abstract}

The intermittency is investigated when the current reversal occurs
in a deterministic inertia ratchet system. To determine which type the intermittency belongs to,
we obtain the return map of velocities of particle using stroboscopic recording, 
and numerically calculate the distribution of average laminar length ${\langle}l{\rangle}$. 
The distribution follows the scaling law of ${\langle}l{\rangle} \propto {\epsilon}^{-1/2}$,
the characteristic relation of type-I intermittency.

\end{abstract}


\maketitle

In the last ten years there has been an increasing interest on the ratchet system. 
The ratchet system has been studied theoretically and 
experimentally in many different fields of science, e.g., molecular motors \cite{Astumian,Bier},
new methods of separation of particles \cite{Rousselet}, 
and current-voltage rectification in asymmetric superconducting 
quantum interference devices\cite{Zapata}. More recently there have been several works 
on ratchet system in quantum domain \cite{Olson,Grifoni}, 
Josephson junction array\cite{Lee}, and etc.

Ratchet system \cite{history} is generally defined as a system 
that is able to transport particles in a periodic structure 
with nonzero macroscopic velocity, although on average 
no macroscopic force is acting \cite{Hanggi}.
For directional motion of particle with no macroscopic force, or 
unbiased fluctuation in a periodic structure, the system has to be 
driven away from thermal equilibrium by an additional 
deterministic \cite{Magnasco,Vicsek} or stochastic pertubation \cite{Bartussek,Dialynas}.
Besides the breaking of thermal equilibrium, for directional motion 
the breaking of spatial inversion symmetry is usually required further \cite{Reimann}.
For this, the potential has a spatially asymmetric shape that is called `ratchet'.
It has also shown that the directional motion can exist 
in the presence of spatially symmetric potential with external pertubation 
that is time-asymmetric \cite{Seeger}.

Many works on ratchet system have been mainly limited on the overdamped case.
In these works, systems in interest are related to microscopic scale 
in which thermal fluctuations or noises play a dominant role.
Recently, Jung, Kissner, and H\"{a}nggi have considered the effect of 
finite inertia on ratchet system \cite{Jung}. 
By considering the inertia term, the dynamics can exhibit both regular and 
chaotic behaviors. They have shown that deterministic chaos, to some extent, mimics
the role of noise. Also, it shows that there exists multiple current reversals as the amplitude of 
external driving is varied.
Thereafter this system has been called `deterministic inertia ratchets' or 
`deterministic underdamped ratchets' \cite{Arizmendi,Barbi}.

In a recent letter, Mateos has shown that the origin of current reversal 
in the deterministic inertia ratchets may be related to a bifurcation 
from chaotic to regular regime \cite{Mateos}.
The observation of intermittent chaos has also been addressed 
before the occurrence of current reversal. However, Mateos has merely mentioned 
the diffusion property in intermittent chaotic regime.
After his work, it has been conjectured that the mechanism of current reversal 
may be related to a crisis bifurcation in which a chaotic state suddenly becomes 
periodic \cite{Arizmendi}.
Furthermore, it was shown that current reversal in the same system can occur even 
in the absence of bifurcation from chaotic to regular regime 
on the other parameter ranges. This current reversal is interpreted 
in terms of change of the stability of the periodic rotating orbits \cite{Barbi}.
At present, up to our knowledge, any further work on the classification of 
the type of intermittency just past the bifurcation has not been reported.

On the other hand, the Pameau and Manneville types of intermittency
are mainly classfied into types I, II, 
and III by the structure of local Poincar\'{e} map, 
$v_{n+1} = v_n + av_n^{2} + \epsilon$, $v_{n+1} = (1 + \epsilon)v_n + av_n^3$, 
and $v_{n+1} = -(1 + \epsilon)v_n - av_n^3$, respectively \cite{Manneville}.
And these types of intermittency are characterized by characteristic relations, 
${\langle}l{\rangle} \propto {\epsilon}^{-1/2}$ for type-I, 
and ${\langle}l{\rangle} \propto {\epsilon}^{-1}$ for type-II and type-III, 
where ${\langle}l{\rangle}$ is the average laminar length. 
Here the parameter $\epsilon$ in type-I intermittency is the channel width 
between the diagonal line and the local Poincar\'{e} map, 
while $1 + \epsilon$ in type-II and type-III is the slope of 
the local Poincar\'{e} map around the tangent point.

The aim of this report is to investigate the intermittency of 
current reversal phenomenon in the deterministic inertia ratchets. 
We take the same system as in Ref. \cite{Mateos}, and explicitly show that 
there exists type-I intermittency when current reversal occurs 
from choatic to regular regime.      

\begin{figure}[htb]
\includegraphics[angle=-90,width=\columnwidth]{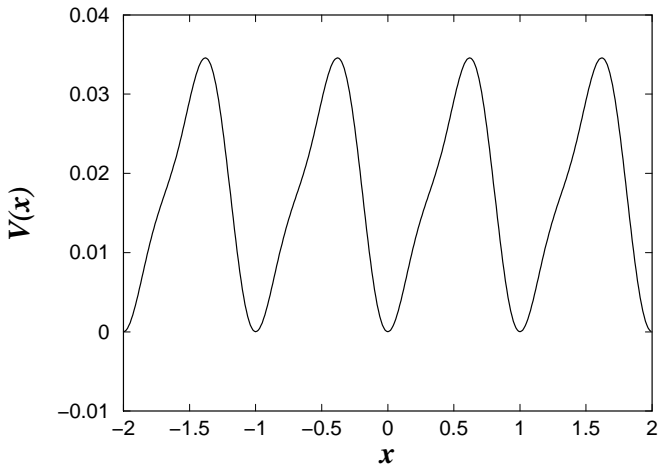}
\caption{The periodic asymmetric potential {\it V}({\it x}) used,
with parameters $C = -(\sin2{\pi}x_{0} + 0.25\sin4{\pi}x_{0})/4{\pi}^{2}\delta$, $x_{0} = -0.19$, 
and $\delta {\simeq} 1.614324$ as in Ref. \cite{Mateos}.}
\label{fig:fig1.eps}
\end{figure}

Now, let us consider a system in which a particle moves in ratchet potential, 
subjected to time-periodic driving and damping. The equation of motion is written as
\begin{equation}\label{eq:1}
\frac{d^{2}x}{dt^{2}} + b\frac{dx}{dt} +\frac{dV}{dx} = a\cos({\omega}t),
\end{equation}
\noindent where b is the friction coefficient, and $\omega$ and {\it a} are
the frequency and the amplitude of the external driving, respectively.
Here $V(x)$ in Fig. \ref{fig:fig1.eps} is the ratchet potential, and is given by
\begin{equation}\label{eq:2}
V(x) = C - \frac{\sin2{\pi}(x-x_{0}) + 0.25\sin4{\pi}(x-x_{0})}{4{\pi}^{2}\delta},
\end{equation}
\noindent where {\it C} and $x_{0}$ are introduced to show 
the potential has minimum at $x=0$ with $V(0)=0$,
and $\delta = \sin(2{\pi}|x_{0}|) + \sin(4{\pi}|x_{0}|)$.  

Among the three dimensionless parameters in Eq.(\ref{eq:1}), 
we vary the parameter $a$, and fix $ b = 0.1 $ and $ \omega = 0.67 $ 
throughout this paper as in Ref. \cite{Mateos}. And
the equation of motion, Eq.(\ref{eq:1}) is nonlinear,
and the inertial term associated with the second derivative of $x$ allows 
the possibility of chaotic orbits. 
We solve this system numerically, using fourth-order Runge-Kutta algorithms.
Because of sensitivity of the above chaotic system, we calculate all values and parameters
including $\delta$ with double precision. 

\begin{figure}[htb]
\includegraphics[angle=-90,width=\columnwidth]{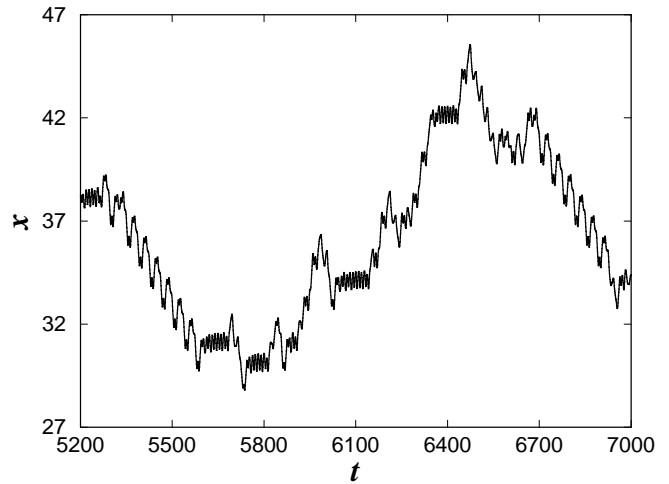}
\caption{For $b = 0.1$ and $\omega = 0.67$, we show the intermittent chaotic trajectory 
of the particle  at $a = 0.080910000$ during current reversal.}
\label{fig:fig2.eps}
\end{figure}

In this system, there exists a bifurcation from chaotic to regular regime 
when current reversal takes place through varying the parameter $a$ \cite{Mateos}. 
As shown in Fig. \ref{fig:fig2.eps}, the particle shows intermittent chaotic behavior 
in the bifurcation region. The particle moves almost regularly 
in a negative direction, and occasionally shows chaotic burst.

\begin{figure}[htb]
\includegraphics[angle=-90,width=\columnwidth]{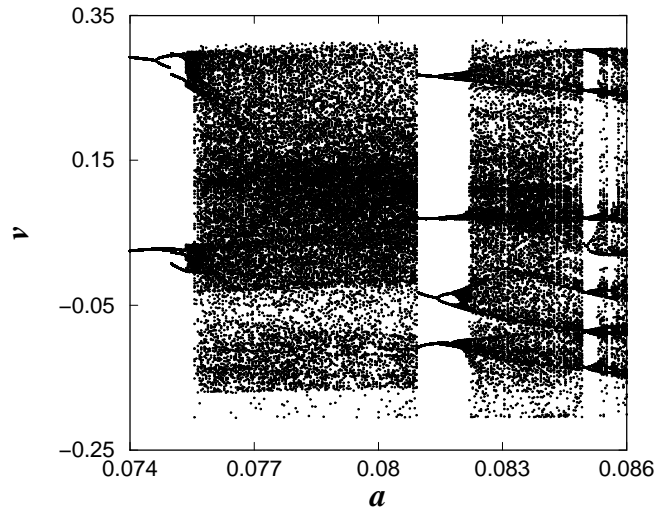}
\caption{The bifurcation diagram with varying the parameter {\it a}, 
other parameters are fixed, $b = 0.1$ and $\omega = 0.67$.}
\label{fig:fig3.eps}
\end{figure}

Furthermore, to study the behavior of the system during the current reversal,
we plot a bifurcation diagram of velocity of a particle. 
For obtaining a bifurcation diagram,
we take stroboscopic recording of the first derivative of {\it x} at times, $t = k\tau$,
where $k$ is the positive integer, and $\tau$ is the period of the external driving, 
$\tau = 2{\pi}/{\omega}$. In the bifurcation diagram, Fig. \ref{fig:fig3.eps}, 
velocities of the particle are plotted on the paramater range of 
$a \in (0.074000000,0.086000000)$,
after long initial transient data are dropped. 
The result is analogous to Fig. 2(a) in Ref. \cite{Mateos}. 

In Fig. \ref{fig:fig3.eps}, two fixed points of velocities plotted 
in the bifurcation diagram at $a = 0.074000000$ correspond to the
{\it regular positive current of particle having period two},
while four fixed points of velocities at $a= 0.080990000$ 
correspond to the {\it regular negative current of particle having period four}\cite{Mateos}.
During this current reversal, there is a period-doubling route to chaos, 
as shown in Fig. \ref{fig:fig3.eps}. 
Bifurcation from chaotic to regular regimes takes place at critical value $a_c$, 
just above $a = 0.080947429$ \cite{comment}. 
After this bifurcation, a periodic window corresponding to the regular negative current 
emerges. 

\begin{figure}[htb]
\includegraphics[angle=-90,width=\columnwidth]{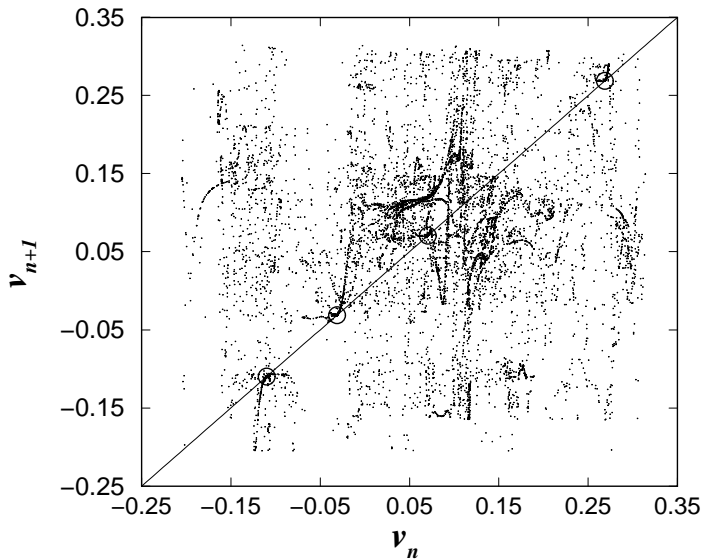}
\caption{The return map, which is plotted at the parameter $a = 0.080947429$ 
just below $a_c$. In the vicinity of $-0.110$, $-0.031$, $0.071$, and $0.270$, 
the return map is nearly tangent to the diagonal line. 
Four open circles indicate nearly tangent regions.}
\label{fig:fig4.eps}
\end{figure}

\begin{figure}[htb]
\includegraphics[angle=-90,width=\columnwidth]{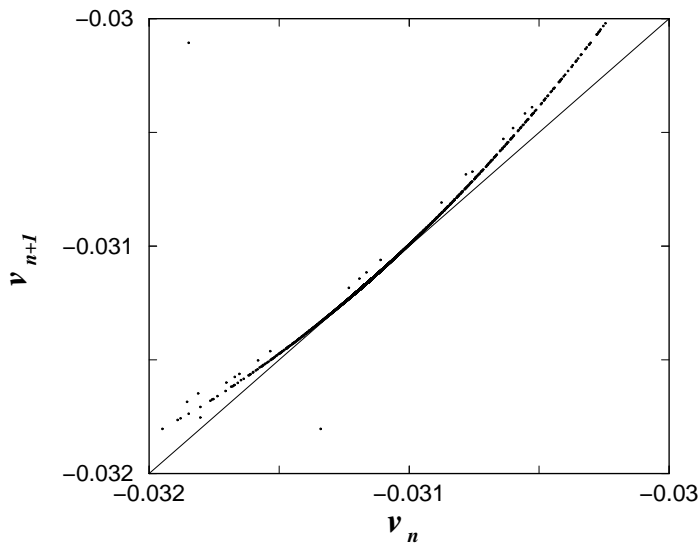}
\caption{The enlargement of Fig. \ref {fig:fig4.eps} in vicinity of $v = -0.031$.
The return map seems nearly tangent to the diagonal line, but it is not exactly tangent.
The channel width between the return map and the diagonal line is in the order of 0.000001.}
\label{fig:fig5.eps}
\end{figure}

To determine which type above intermittency belongs to,
we numerically obtain the return map, $f^{4}(v_n)$,
which shows the relation between the velocity of particle, $v_n$ and 
the velocity of particle, $v_{n+1}$ after $4\tau$ time interval elapse, 
with external driving period $\tau$, $\tau = 2{\pi}/{\omega}$.
We use the value of $\tau$ with double precision because of the sensitivity of the
chaotic system. 
In Fig. \ref{fig:fig4.eps}, we plot the return map at just below $a_c$. 
Fig. \ref{fig:fig5.eps} is enlargement of the part of Fig. \ref{fig:fig4.eps} 
in the vicinity of $v_{n}=-0.031$.
The points on the digonal line in the return map correspond to 
the states of $v_{n+1} = v_n$, i.e, four periodic motion.


\begin{figure}[htb]
\includegraphics[angle=-90,width=\columnwidth]{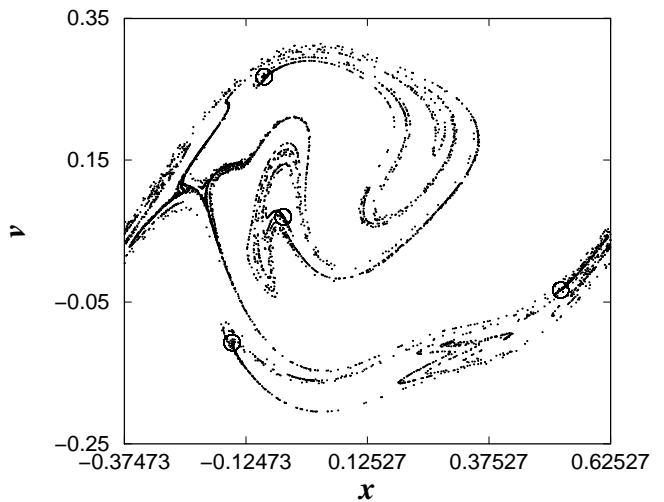}
\caption{The phase portrait of two attractors of the ratchet equation 
for $b = 0.1$ and $\omega = 0.67$:
one is the chaotic attractor for $a = 0.080947429$, 
just below $a_c$ and the other is the period four attractor
for $a = 0.080990000$, represented by the center of four open circles.}
\label{fig:fig6.eps}
\end{figure}

As shown in Fig. \ref{fig:fig5.eps}, the return map is nearly tangent to 
the diagonal line at the parameter $a$ just below $a_c$. 
When a particle remains in a narrow channel, the trajectory shows 
almost a regular periodic behavior. After the particle escapes 
from this narrow channel, the trajectory shows a chaotic burst, and 
then goes back to the channel. (This process keeps repeating.)
As increasing $a$ from the value of just below $a_c$ to $a_c$, 
the channel width narrows more and more so that the particle spends most of the time in the channel.  
After all, the return map crosses the diagonal line, 
when $a$ becomes larger than $a_c$, and two crossing points 
corresponding to the stable and unstable fixed points are made. 
Among these points, the stable fixed point corresponds to the
regular negative current of particle.
Like this, the intermittency emerges, before the return map undergoes a tangent bifurcation.
This analysis well agrees with the result shown in Fig. \ref{fig:fig6.eps}, 
which is similar to that of Mateos\cite{Mateos}.

In Fig. \ref{fig:fig6.eps},  
one of two attractors is the chaotic one for $a = 0.080947429$, 
just below $a_c$, and the other is the period four attractor 
for $a = 0.080990000$, corresponding to the regular negative current.
Note that these two attractors are obtained by confining the dynamics 
in one potential well that include $x = 0$, i.e,
in the range of $x \in (-0.37473443,0.62526557)$.
The latter consists of four points at the phase space.
In the chaotic attractor, a particle spends most of the time in the vicinity of 
four point attractors corresponding to the regular negative current, 
and once in a while intermittently moves in a chaotic way.

\begin{figure}[htb]
\includegraphics[angle=-90,width=\columnwidth]{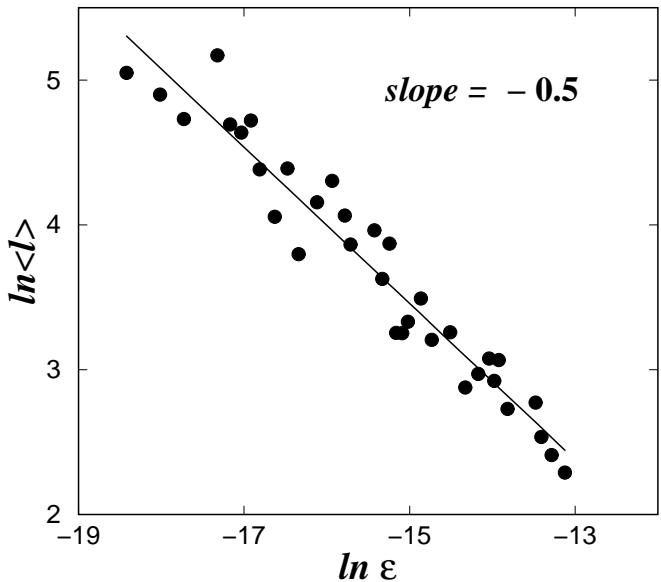}
\caption{The distribution of the average laminar length with varying the parameter {\it a}. 
It follows the scaling law of ${\langle}l{\rangle} \propto {\epsilon}^{-1/2}$ 
with $\epsilon = a-a_c$.}
\label{fig:fig7.eps}
\end{figure}

So far, we have investigated the type of intermittency qualitatively.
For quantitative characterization of the type of intermittency, let us survey 
the scaling law for duration of the laminar state, 
which means the particle remains in the narrow channel, 
as shown in Fig. \ref{fig:fig5.eps}.
We calculate the average laminar length ${\langle}l{\rangle}$ by averaging 
the iterated numbers of the return map at the laminar state, 
as varying $a$ from the range of below $a_c$ to $a_c$.
Numerically, we take the laminar state is the situation 
that the difference between $v_{n+1}$ and $v$ is less than $0.0003$. 
The result has shown in Fig. \ref{fig:fig7.eps}.
The distribution of the average laminar length follows the scaling law of 
${\langle}l{\rangle} \propto {\epsilon}^{-1/2}$ with $\epsilon = a-a_{c}$.
It agrees with that the distribution of the average laminar length 
of the type-I intremittency is of the form 
${\langle}l{\rangle} \propto {\epsilon}^{-1/2}$ 
if there are no external noises \cite{Manneville,Kim,Kye}.
Therefore, the intermittency that exists before the bifurcation taking place 
from the chaotic to the regular regime is the type-I intermittency.    

In conclusion, we have investigated the mechanism of current reversal 
in deterministic inertia ratchets. By numerically obtaining the return map 
of velocities, and the scaling law giving the characteristic relation of intermittency,
we have explicitly shown that the type-I intermittency exists 
when the current reversal occurs from the chaotic to the regular regime. 

The authors thank D.-U. Hwang, J.-W. Ryu, and M.-U. Kim for valuable discussions.
This work was supported by the Creative Research Initiatives of the
Korean Ministry of Science and Technology. Two of us(W.-S.S and Y.-J.P.) 
acknowledge support from the Sogang University Research Grants in 2003.


\begin{thebibliography}{99}
\bibitem{history}In analogy with the ratchet system considered by 
{\em The Feynman Lectures on Physics}, Vol. I, Chap. 46 
in discussing the devices that are in contradiction with 
the second law of thermodynamics, these systems are also called ratchets.
\bibitem{comment} In Ref. \cite{Mateos}, Mateos showed the critical parameter value,
$a_c {\simeq} 0.08092844$ by using $\delta {\simeq} 1.6$.
The difference with our result is due to the parameter value $\delta$.
We use $\delta = 1.61432359181694$ with double precision,
for satisfactory results on the return map of velocities in Fig. \ref{fig:fig5.eps}
and the distribution of average laminar length in Fig. \ref{fig:fig7.eps}.
\end{thebibliography}
\end{document}